\documentclass[ ]{article}
\usepackage{physics}
\usepackage{float}
\usepackage{graphicx}
\usepackage[margin=0.8 in]{geometry}
\usepackage{amsmath}
\usepackage{booktabs}
\usepackage{authblk}
\usepackage{lscape}
\usepackage{bbm}
\usepackage{hyperref}

\author[\#]{Ruchira Mishra\footnote{email: ruchiramishra98@gmail.com, ms16071@iisermohali.ac.in}}
\affil[\#]{\it Indian Institute of Science Education and Research Mohali 140306, India}
\author[\#]{Andrea Vinante\footnote{email: A.Vinante@soton.ac.uk}}
\affil[\#]{\it Physics and Astronomy, University of Southampton, Southampton SO17 1BJ, U. K.
}
\author[3]{Tejinder P. Singh\footnote{email: tpsingh@tifr.res.in}}
\affil[3]{\it Tata Institute of Fundamental Research, Homi Bhabha Road, Mumbai 400005, India}

\title{Testing Spontaneous Collapse Through Bulk Heating Experiments\\ - {\it Estimate of the Background Noise - }}

\begin{document}

\maketitle


\begin{abstract}
\noindent Models of spontaneous wave function collapse  predict a small heating rate for a bulk solid, as a result of coupling to the noise field that causes collapse. This rate is small enough that ambient radioactivity and cosmic ray flux on the surface of the earth can mask the heating due to spontaneous collapse. In this paper we estimate the background noise due to gamma-radiation and cosmic ray muon flux, at different depths. We demonstrate that a low-temperature underground experiment at a depth of about 6.5 km.w.e. would have a low enough background to allow  detection of bulk heating for a collapse rate $\lambda$ of $10^{-16}$ s$^{-1}$ using presently available technology.
\end{abstract}

\section{Introduction}
Models of spontaneous wave function collapse \cite{GRW, GRWP} provide a falsifiable solution to the quantum measurement problem, by making a stochastic nonlinear modification of the Schr\"odinger equation [for reviews see e.g. \cite{BG, RMP}]. A hypothesised  noise field is responsible for the collapse of the wave function, and one side effect of the noise field is that it causes the collapsed object to gain energy from the noise field at a rate given by \cite{b,b2}
\begin{equation}\label{key}
\derivative{E}{t}=\frac{3}{4}\lambda\frac{\hbar^2}{{r_c}^2}\frac{M}{{m_N}^2}
\end{equation}
Here $M$ is the mass of the collapsed object, $m_N $ is the nucleon mass, $ \lambda$ the collapse rate and coupling parameter with noise field of correlation length $ r_c $. The original GRW collapse model, and the currently favoured best-studied CSL collapse model both propose a value for the collapse rate $\lambda_{GRW}=\lambda_{CSL}= 10^{-16}$ s$^{-1}$ and $r_c=10^{-7}$ m. Eventually however, the values of these parameters must be determined by experiments. Taking the preceding value for $r_c$ and keeping $\lambda$ a free parameter, the heating rate per unit mass, $H\equiv dE/Mdt$ is hence given by
\begin{equation}
H \simeq 0.32\; \lambda  \; {\rm J/kg}
\label{heatingrate}
\end{equation}
The experimental attempts and proposals to detect this heating, or the random motion induced by this heating, constitute the 
non-interferometric tests of collapse models, and currently provide the most stringent bounds on the values of $\lambda$ and $r_c$. Current experimental evidence based on mechanical experiments such as ultracold microcantilevers \cite{vinante1,vinante2} or gravitational wave detectors \cite{carlessoLISA, LISA2} puts an upper bound such that $\lambda \leq 10^{-8}$ s$^{-1}$ at $r_c=10^{-7}$ m, implying that the heating rate is smaller than about $10^{-9}$ W/kg. A stronger bound at level $\lambda \leq 10^{-11}$ s$^{-1}$ is set by spontaneous emission of x-rays from free electrons in solid germanium \cite{curceanu}, although this technique relies on the assumption that the CSL noise is white up to the frequency of x-rays. 

Detection of bulk CSL heating of a solid through coupling of its phonon vibrations to the collapse noise field is another promising way to test collapse models, when the experiment is done at ultra-low temperatures (about a milli-kelvin) \cite{a,Bahrami}. However, 
background noise is an important factor to be accounted for while doing these experiments - either by making sure that heating due to these effects is negligible as compared to CSL heating or by looking for unique signatures of these effects. In particular, cosmic ray muons and ambient radioactive decays provide a heating rate of about $10^{-10}$ W/kg, and this noise must be carefully accounted for or reduced/eliminated, if the CSL heating has to be confirmed or ruled out. In this paper we compare the magnitude of CSL heating with this background noise, and estimate the amount of rock cover needed to sufficiently cut out the muon flux, so that the value $\lambda_{CSL}\simeq 10^{-16}$ s$^{-1}$ becomes detectable. We demonstrate that this value is detectable with current technology at a depth of  6.5 km.w.e. Currently, the world's deepest laboratory for physics experiments is the Jinping Underground Laboratory in China, at a depth of 6.7 km.w.e. \cite{china}. [Kilometer water equivalent is a standard measure of cosmic ray attenuation in underground laboratories. A laboratory at a depth of 1 km.w.e is shielded from cosmic rays equivalent to a lab 1 km below the surface of a body of water. Because laboratories at the same depth (in kilometers) can have greatly varied levels of cosmic ray penetration, the km.w.e. provides a convenient and consistent way of comparing cosmic ray levels in different underground locations.]

Just as for x-ray techniques, bulk heating experiments are also sensitive to specific spectral components of the CSL field, with relevant frequency of the order of $10^{11}$ Hz for typical solid matter \cite{a}. However, if CSL is a cosmological field, this frequency seems quite more realistic than in the x-ray case, as most known cosmological fields (such as CMB or relic neutrinos) extend up to $10^{11}$-$10^{12}$ Hz.

\section{Experimental setup: the need to go underground}

The experimental setup proposed in \cite{a} is shown in Fig. 1 and is essentially an ultrasensitive low temperature bolometer consisting of three components - CSL noise absorber of mass $m$, resistor with thermal resistance $R$, and the thermal bath, characterized by a constant temperature $T_0$. The absorber and thermal bath are connected through the resistor.
\begin{figure}[H]
	\centering
	\includegraphics[width=0.7\linewidth]{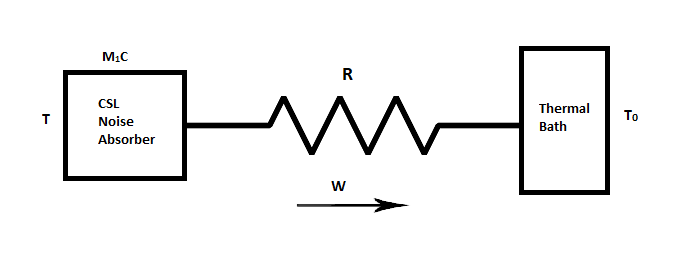}
	\caption{Schematic diagram of the experimental set-up}
\end{figure}
If the experiment is performed at ultra-low temperatures of the order of  millikelvin, the specific heat $c$ and hence the heat capacity $mc$ are extremely low, while thermal resistances can be very high. This makes the system extremely sensitive to any input power.

Clearly, the system will be also extremely sensitive to any background event, such as the absorption of gammas or cosmic muons, and the obvious way to minimize the background heating rate is to go underground. Background events could  also be detected and their effect subtracted, provided that they are sufficiently rare. For an impulsive event of energy $E$, a sharp rise of temperature with peak height $E/mc$ arise, followed by an exponential decay with time constant $ \tau= {mcR} $. In contrast, for continuous power deposition $W$, as due to CSL heating, a constant temperature gradient is expected $T-T_0=R W$. So, in order to have a large temperature gradient, $R$ has to be kept large.

Assuming background is completely suppressed, let us estimate the effective CSL-induced temperature gradient expected in a realistic experiment. For concreteness, consider as reference the CUORE experiment, looking for neutrinoless double beta decay \cite{CUORE0}. It is based on cubic masses made of TeO$_2$ with side $5$ cm, mass $m=750$ g, cooled down to a temperature $T=10$ mK. Heat capacity is $\simeq 2 $ nJ/K while thermal resistance, provided by a PTFE support structure, is $R \simeq 2 \times 10^8$ K/W, resulting in a time constant $\tau \simeq 0.5$ s. 

With such values the CSL-induced temperature increase would be of the order of $5$ mK for $ \lambda \simeq 10^{-10}$ Hz, which is a relatively large and measurable effect. The temperature resolution of such experiment is actually much better, the temperature being measured by high-responsivity low noise thermistors. However, the issue of an experiment looking for CSL is that the heating is constant and cannot be modulated or turned off and on. Therefore, the relevant experimental parameter is a good absolute temperature accuracy, rather than a high temperature sensitivity. Relative temperature accuracy better than $5 \%$ can be reached for instance by noise thermometry \cite{enss}, allowing $\lambda \simeq 10^{-11}$ Hz to be unambiguously probed in a CUORE-like experiment. In order to explore the parameter space down to the standard CSL collapse strength $\lambda=10^{-16}$ Hz, one can in principle increase the mass by a factor $10$ and reduce the temperature to about $1$ mK. This would both increase the thermal resistance by a factor of roughly $10^3$ (contact thermal resistance usually scales as $T^{-3}$) and improve the absolute temperature accuracy by a factor $10$. 

Thermodynamic temperature fluctuations, a typical limiting factor of ultracold thermal detectors such as bolometers or cryogenic microcalorimeters, will be largely negligible in the proposed experiment. In fact, temperature fluctuations are of the order of $\sqrt{k_B T^2/C_v}$ (see for instance Landau-Lifshitz, Physical Statisics, Chapter XII) where $T$ is the temperature, $k_B$ is Boltzmann constant and $C_v$ is the heat capacity. Conventional bolometers/calorimeters are designed to have very low heat capacity, in order to resolve single photons or detect very low microwave flux, and typically approach this fundamental limit. In contrast, due to much larger mass and heat capacity, this is not going to become an issue for the proposed experiment.

Eventually, the main limiting factor is constituted by background particles events. Heating due to highly penetrating background particles such as cosmic muons scales roughly with the volume of the test mass, similarly as the CSL effect. This is because the particle flux goes with area and the deposited energy per event with thickness. Therefore, background particles will impose a fundamental limit to this type of experiment in absence of proper shielding.

Furthermore, an experiment looking for very small heating will necessarily have long time constants. For a configuration similar to CUORE the time constant is of the order of 1 s, but for a larger mass/lower temperature experiment the time constant will likely increase by 1-2 orders of magnitude. This calls for an extremely low background rate, if one wishes to resolve single events and subtract their effect. 

These considerations imply that a CSL bulk heating experiment has to be performed underground. Since the frequency of the background events reduces as we go deeper \cite{c}, we can afford to have larger time constant and hence high sensitivity underground, something that is not possible on the surface of the earth. In the following two sections we will estimate the mean absorbed power due to background $\gamma$-radiation and cosmic ray muons. For concreteness, we choose Germanium as reference material for the CSL noise absorber, as its properties are well characterized. We take its density to be $ \rho=5.67 \;\text{g{cm}}^{-3} $.

\section{Effect of $ \gamma $ radiation}
Gamma radiation results from the background as well as the materials surrounding the detector. Assuming that ultra-pure materials with negligible amount of impurities giving rise to radioactivity are used, we estimate the variation in power deposited as a function of shield thickness, for a Ge detector  with side $l=10$ cm. The gamma ray flux at LNGS (Gran Sasso) was obtained from \cite{i}.

\subsection{Calculating the power deposited by $ \gamma $ radiation}
The interaction of a photon with any material depends on its photon scattering cross section which is dependent on its energy and the properties of the interacting material.The fraction of incident photons transmitted through a material, given by the transmission factor s, is a function of this photon scattering cross section, density of the material and the intervening length. The transmission factor $s$ is given by
 \begin{equation}\label{key}
s=\exp(-{({\mu}/{\rho})}_{tot,t,x}\rho_x l_x)
\end{equation} where ${({\mu}/{\rho})}_{tot,t,x}$ is the photon scattering cross section \cite{j} for a photon with a given energy interacting with a material $ x $ of density $ \rho_x $ and intervening length $ l_x $. For a large shield (or  low energy photons), most of the photons are absorbed and hence result in a very small transmission factor.
For a detector kept inside a shield, a very small fraction of the incident photons will be transmitted to the detector. Further, only a fraction of these transmitted photons will be absorbed by the detector. This fraction is given by
\begin{equation}\label{key}
p=1-\exp(-{({\mu}/{\rho})}_{tot,t,d}\rho_d l_d)
\end{equation}
 where ${({\mu}/{\rho})}_{tot,t,d}$ is the photon cross scattering cross section for the photons incident on a detector with density $\rho_d  $ and intervening length $l_d  $.
Furthermore, only a fraction of the absorbed photons deposit energy to the detector. This fraction is  obtained using the energy absorption coefficient \cite{d}  \begin{equation}\label{key}
E_{abs}/E_{\gamma}=1-\exp(-{({\mu}/{\rho})}_{tot,en,d}\rho_d l_d)
\end{equation} 
where ${(\mu}/{\rho)}_{tot,en,d}$ is the energy absorption coefficient \cite{j} for a photon with a specific energy incident on  detector $ d $.
Then, in order to calculate the power deposited to the detector, the incident $\gamma$ spectrum energy is multiplied with the flux, total incident area of the detector, transmission factor $s$ of the shield, absorption factor of the detector and the energy absorption coefficient of the detector \cite{d}.
Thus, the power deposited to the detector is given by, \begin{equation}\label{key}
P=E\times J_\gamma\times A\times s_{shield} \times p_{detector}\times E_{abs}/E_{\gamma}
\end{equation}
where $ E $ is the incident $\gamma$ spectrum energy, $ J_\gamma $ the flux, $ A $ is total incident area of the detector,$ s_{shield} $ is transmission factor of the shield, $ p_{detector} $ is absorption factor of the detector and $E_{abs}/E_{\gamma}  $ is the energy absorption coefficient of the detector \cite{d}.
Integrating this value over all energies gives the total power deposited.

\subsection{Calculating Power deposited as a function of shield thickness}
Lead (Pb) having a higher photon  scattering cross-section, absorbs most of the radiation resulting in very small $s$ values. Hence, it is preferred for shielding $ \gamma $ radiation. For a Ge cube with l=10 cm, power deposited as a function of shield thickness is plotted in Fig. 2. 

The values of gamma ray flux at LNGS \cite{i} were used for calculations. Data from \cite{j} was used for calculating the values of $ s $ and $ p $ for the shield and detector. The values of power deposited thus obtained were extrapolated for the entire range of 0-3 MeV, using a linear fit for the ranges 0-0.5 MeV, 0.5-1 MeV, 1-2 MeV and 2-3 MeV. Further the total power deposited was calculated by integrating over these intervals. This was done for different thickness of shields. The error involved in fitting the data is shown. Error in flux values from \cite{i} have not been taken into account.  
\begin{figure}[H]
	\centering
	\includegraphics[width=1\linewidth]{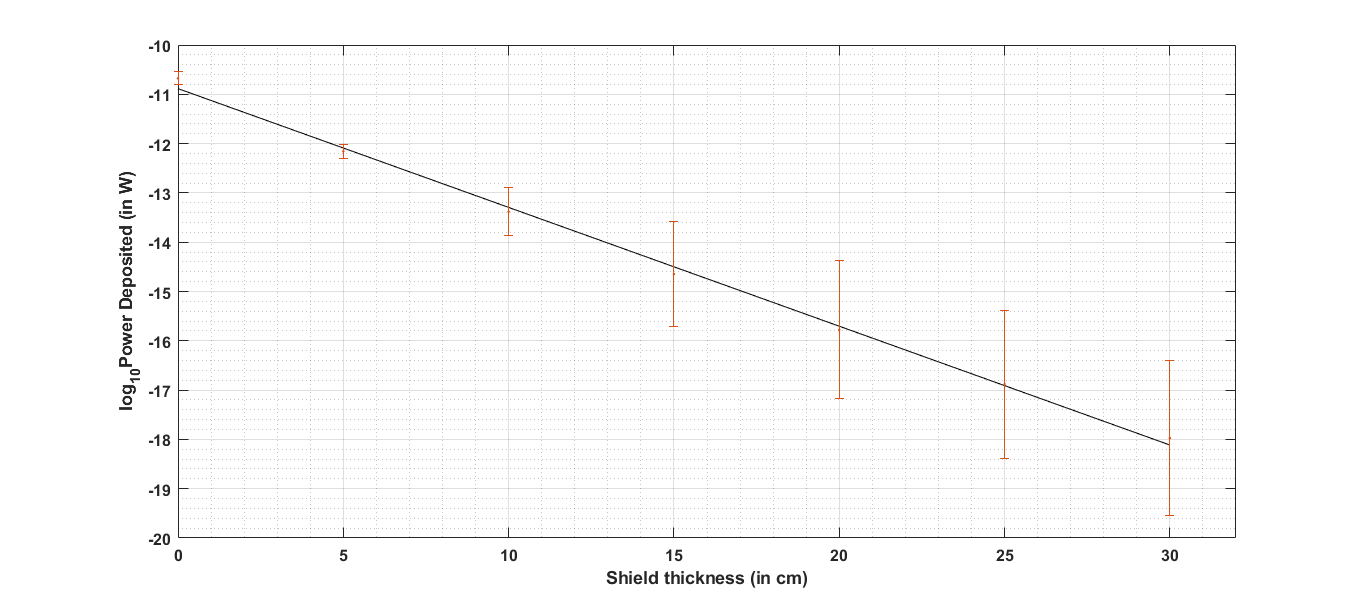}
	\caption{Power deposited (on a Ge cube with l=10cm) as a function of Pb shield thickness. The power deposited varies as $ \log_{10} P=-0.24t-10.9$ where $P$ is the power deposited and $t$ is the thickness of the shield. Value of incident flux  from \cite{i} has been used.}
\end{figure}
For these calculations, it was assumed that the lead being used for shielding is ultra-pure and does not contribute to radioactivity. This might not be a very realistic assumption to make. But, if the contribution from shield was to change the power deposited even by a few orders of magnitude, these events would be very rare to affect our signal \cite{i} even  for a relatively large time constant. So, signal for CSL heating could be obtained after post-processing analysis.

 A similar analysis can also be done for background neutrons, which are shielded using hydrogen rich materials which reduce the number of events.
 It can thus be shown that gamma radiation and neutrons can be
 completely shielded, and can be neglected when looking for CSL heating effects. The important factor to take care of is cosmic ray muons, as we discuss below.

\section{Effect of Cosmic rays}
The most abundant components of cosmic rays present underground are muons, apart from secondaries - gamma rays and neutrons. Muons can easily cross large amounts of matter and penetrate to underground laboratories. Since they cannot be shielded out completely, the only solution is to go deeper in order to reduce their effect.
\subsection{Calculating the power deposited by muons} 
Heating of a detector due to cosmic rays can be calculated using the Bethe-Bloch formula\begin{equation}\label{key}
\derivative{E}{(\rho x)}=\frac{2Cm_ec^2}{\beta^2}\left [\ln\left(\frac{2m_ec^2{E'}_m\beta^2}{(1-\beta^2)I^2}\right)-2\beta^2\right ]\; \text{MeV{cm}}^2\text{g}^{-1}
\end{equation} where,
\begin{equation}\label{key}
C=\frac{\pi{r_0}^2N_AZ}{A}
\end{equation}  
 \begin{equation}\label{avgE}
{E'}_m=2m_ec^2\frac{{p_\mu}^2c^2}{{(m_ec^2)}^2+{(m_\mu c^2)}^2+2m_ec^2\sqrt{{p_\mu}^2c^2+{(m_\mu c^2)}^2}} 
\end{equation}
 \begin{equation}\label{tmoo}
 p_\mu c=\frac{m_\mu c^2\beta}{\sqrt{1-\beta^2}}=\sqrt{{(T_\mu +m_\mu c^2)}^2-{(m_\mu c^2)}^2} 
 \end{equation} and  \begin{equation}\label{key}
 \beta=v/c 
 \end{equation}
\vspace{2mm}\newline In the above equations, $ r_0=2.82\times {10}^{-13} $ cm is the classical radius of the electron, $ N_A $ is Avogadro's number, $ Z $ the atomic number, $ A $ the atomic mass, $ m_ec^2=0.511$ MeV  and $ m_\mu c^2=106$ MeV. $ T_\mu $ is the kinetic energy of the muon ($ \mu $) in MeV and $ v $ is its speed. $ I $ is the mean excitation potential given by $(10eV)\times{Z}$. In the equation above we have neglected higher order corrections to the Bethe-Bloch formula.

If the total muon flux is $ I_v $ in units of $ \text{{cm}}^{-2}\text {s}^{-1}\text{{sterad}}^{-1} $, then directional flux is given by $ I(\theta)=I_v{cos}^2\theta $ where $ \theta $ is the angle between the normal to the horizontal surface and the direction of the incoming rays. Integrating over all incident angles gives the horizontal flux in $\text{{cm}}^{-2}\text{s}^{-1} $\begin{equation}\label{key}
J_1=\frac{\pi}{2}I_v
\end{equation}
If the unit area is tilted at an angle $ \theta_a $ to the vertical, $ J_{\theta_a} $ is given by the expression\begin{equation}\label{key}
J_{\theta_a}=\frac{\pi}{2}I_v(\cos\theta_a+\frac{\pi}{4}\sin\theta_a)
\end{equation}
So, the flux through a vertically oriented surface i.e. $ \theta_a=\frac{\pi}{2} $ will be given by\begin{equation}\label{key}
J_3=\frac{\pi^2}{8}I_v
\end{equation}
Using these values of flux, the total number of events $ \text{s}^{-1} $ can be calculated.
\begin{equation}\label{key0}
\text{No. of events {s}}^{-1}=J_1(\text{horizontal area})+J_3(\text{vertical area})
\end{equation}
The total power then deposited will be given by \cite{d}\begin{equation}\label{key}
P=\derivative{E}{(\rho x)}\times ({\text{no. of events}})\ \text{s}^{-1}\times \rho\times l
\end{equation}
where $ \rho $ is the density of the material and $ l $ is the mean length travelled by the muon. The latter can be obtained through Monte-Carlo simulations. For simplicity we take as approximate value for $l$ the actual side of the cubic absorber.

The variation of vertical muon intensity $ J_3 $ with depth for standard rock (A=22, Z=11, $ \rho=2.65\; \text{g{cm}}^{-3} $) and Gran-Sasso was obtained from \cite{e} and their average energy for a depth $d$, to be used in (\ref{tmoo}), was calculated using \cite{f}\begin{equation}\label{key}
\expval{E_{\mu}}=\frac{\epsilon_{\mu}(1-e^{-\text{bd}})}{\gamma_{\mu}-2}
\end{equation}
Parameters given by \cite{g} (b=$ 0.383/\text{km.w.e}, \gamma_{\mu}=3.7$ and $ \epsilon_{\mu}=618$ GeV) and \cite{h} (b=$ 0.4/\text{km.w.e}, \gamma_{\mu}=3.77$ and $ \epsilon_{\mu}=693$ GeV) for standard rock were used. The energy values thus obtained were assumed to be the same for Gran-Sasso. 
The event rate is obtained from Eqn. (\ref{key0}). The event rate and power deposited was then calculated for a Ge cube of side 10 cm. The resulting values are shown in Figs. 3, 4 and Figs. 5, 6 respectively.
\begin{figure}[H]
	\centering
	\includegraphics[width=1\linewidth]{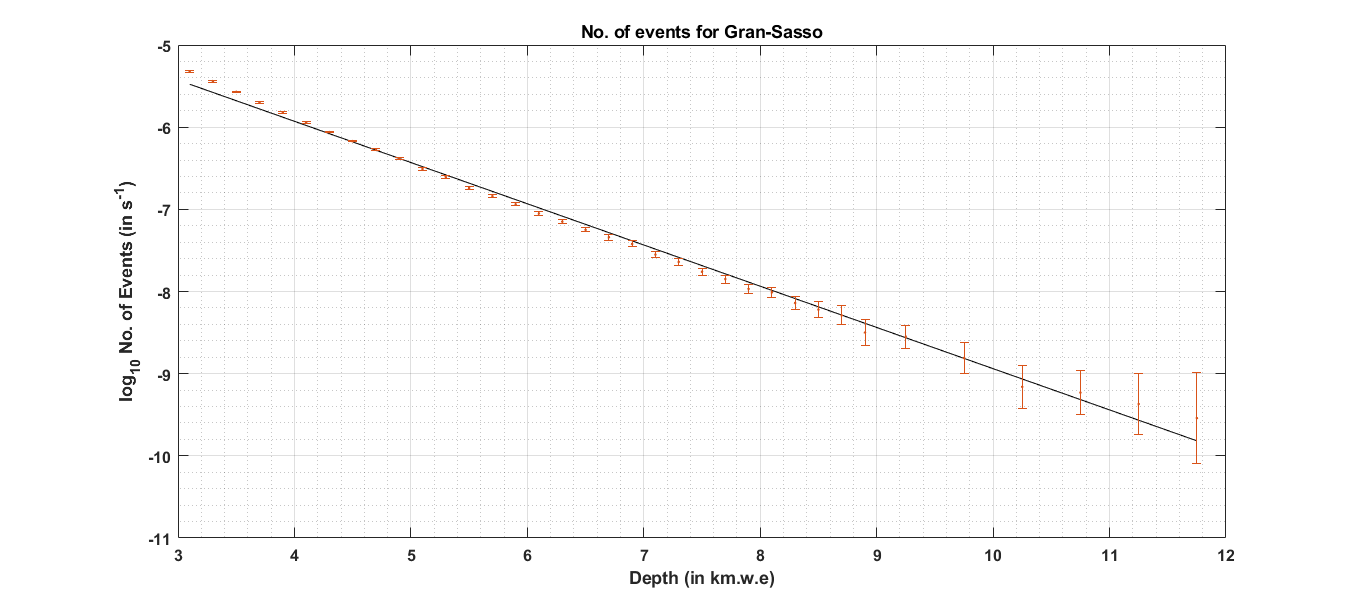}
	\caption{Variation of no. of events of cosmic ray muons with depth for Gran-Sasso (GS) for a 10 cm cube. The event rate is obtained from Eqn. (\ref{key0}). The GS line is fitted to
	the function $y= -0.50x - 3.92$. The intensity values have 
	been taken from [19] and errors in them have been taken into account.}
\end{figure}
\begin{figure}[H]
	\centering
	\includegraphics[width=1\linewidth]{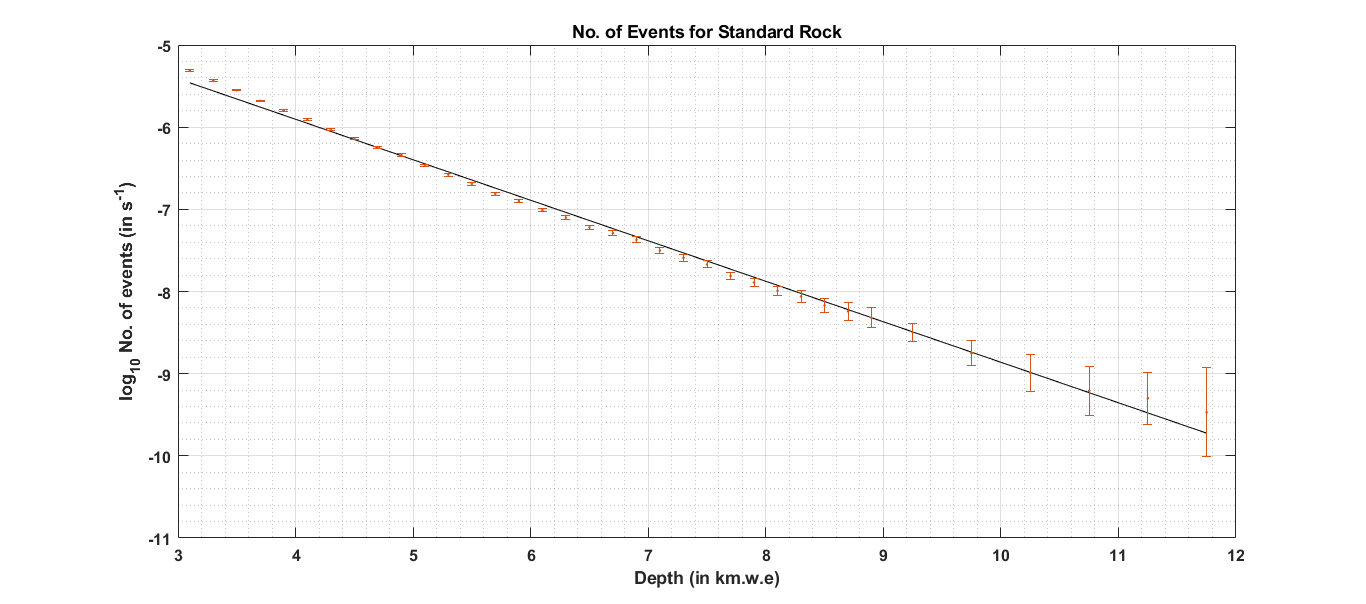}
	\caption{Variation of no. of events of cosmic ray muons with depth for  Standard Rock (A=22, Z=11, $ \rho=2.65\; \text{g{cm}}^{-3} $) (SR) for a 10 cm cube. The event rate is obtained from Eqn. (\ref{key0}). The SR line is fitted  to $y= -0.49x - 3.93$. The intensity values have 
	been taken from [19] and errors in them have been taken into account.}
\end{figure}
\begin{figure}[H]
	\centering
	\includegraphics[width=1\linewidth]{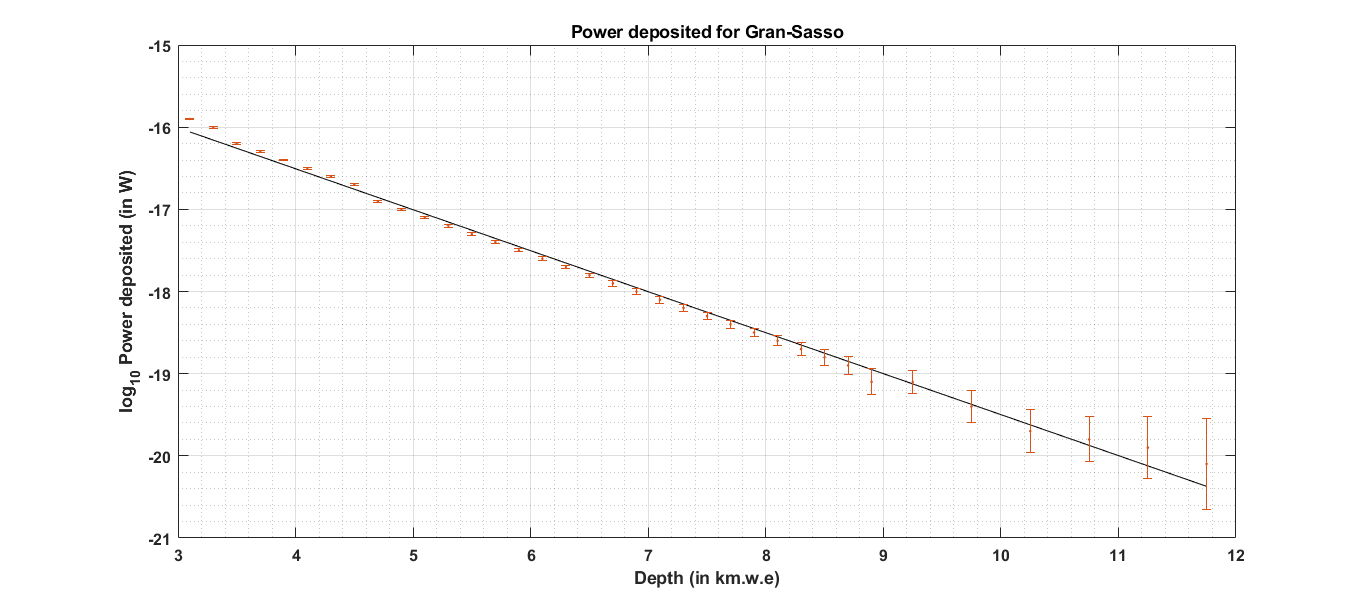}
	\caption{Power deposited  in cosmic ray muons,  as a function of depth. These values are obtained from
	Eqn. (16). The GR line is fitted to  GS  $y= -0.50x -14.51$.  The intensity values have 
	been taken from [19] and errors in them have been taken into account. An error of 4\% has been assumed in the parameters for average energy taken from [21] and [22]. } 
\end{figure}
\begin{figure}[H]
	\centering
	\includegraphics[width=1\linewidth]{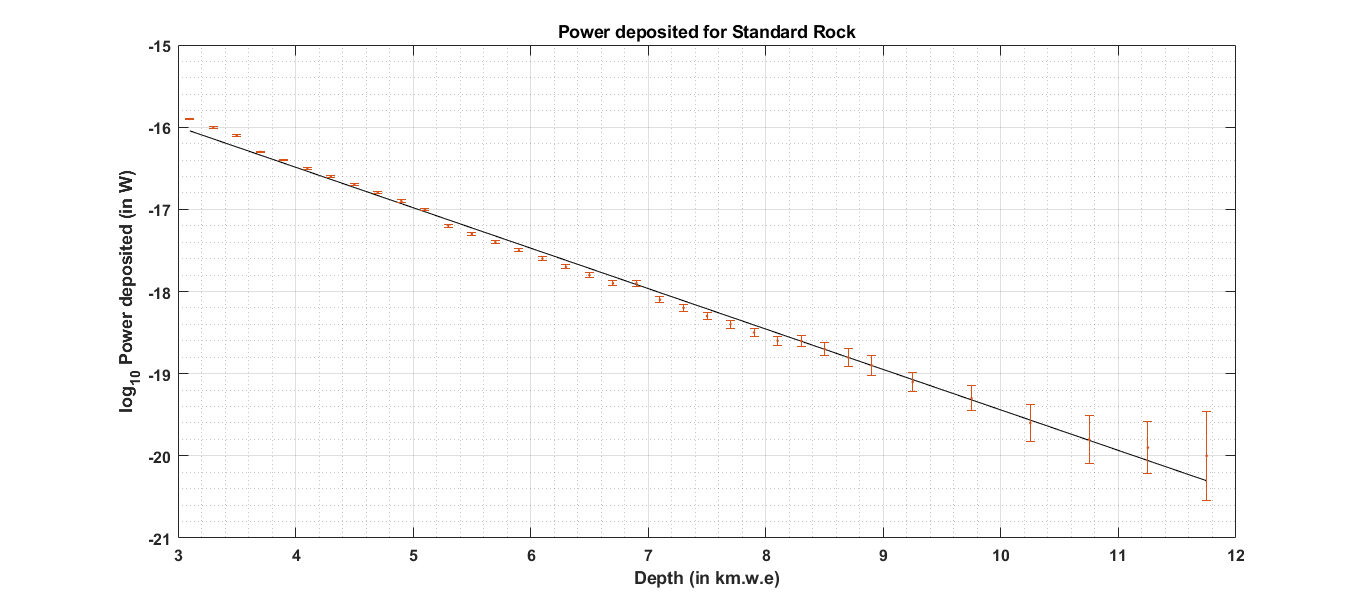}
	\caption{Power deposited  in cosmic ray muons,  as a function of depth. These values are obtained from
	Eqn. (16). The SR line  is fitted to $y= -0.49x - 14.52$. The intensity values have 
	been taken from [19] and errors in them have been taken into account. An error of 4\% has been assumed in the parameters for average energy taken from [21] and [22]. } 
\end{figure}

\subsection{Calculating the value of detectable $\lambda$  as a function of depth}
To get an idea of detectable $ \lambda $ as we go deeper, power deposited by muons at different depths was calculated using the above formulae. A Ge cube with $l=10$ cm was used for the calculations. Assuming that a difference of two orders of magnitude between CSL heating and power deposited by muons is good enough to neglect the effect of muons, the value of $ \lambda $ was estimated for $ r_c={10}^{-7}\; \text{m} $.
\begin{equation}\label{key}
100\times P=\derivative{E}{t}\times\rho\times\text{V}
\end{equation}
Here $ \derivative{E}{t} $ is obtained from (1), $ \rho $ is the density of the material and $V$ its volume.

It is important to note that the variation of intensity becomes non-linear as we go towards the surface. This is because only the most energetic muons are able to penetrate through, thus leaving a large fraction of less-energetic ones behind. This non-linearity is also visible in the values of $ \lambda $ as a result. From extrapolation of the graphs, we expect to detect a value of $ \lambda \approx 10^{-13} \text{s}^{-1} $ on the surface. This of course is not true. On the other hand, using the value of intensity of cosmic ray muons detected on the surface, $ I=1.14\times 10^{-2} \text{cm}^{-2} \text{s}^{-1} \text{sterad}^{-1} $ \cite{z} with average energy 4GeV, gives a value of $ \lambda\approx10^{-9}\text{s}^{-1} $. This is in agreement with values obtained from other experiments.
Figs. 7 and 8 show the value of detectable $\lambda$ as a function of depth, for $r_c=10^{-7}$ m.
\begin{figure}[H]
	\centering
	\includegraphics[width=1\linewidth]{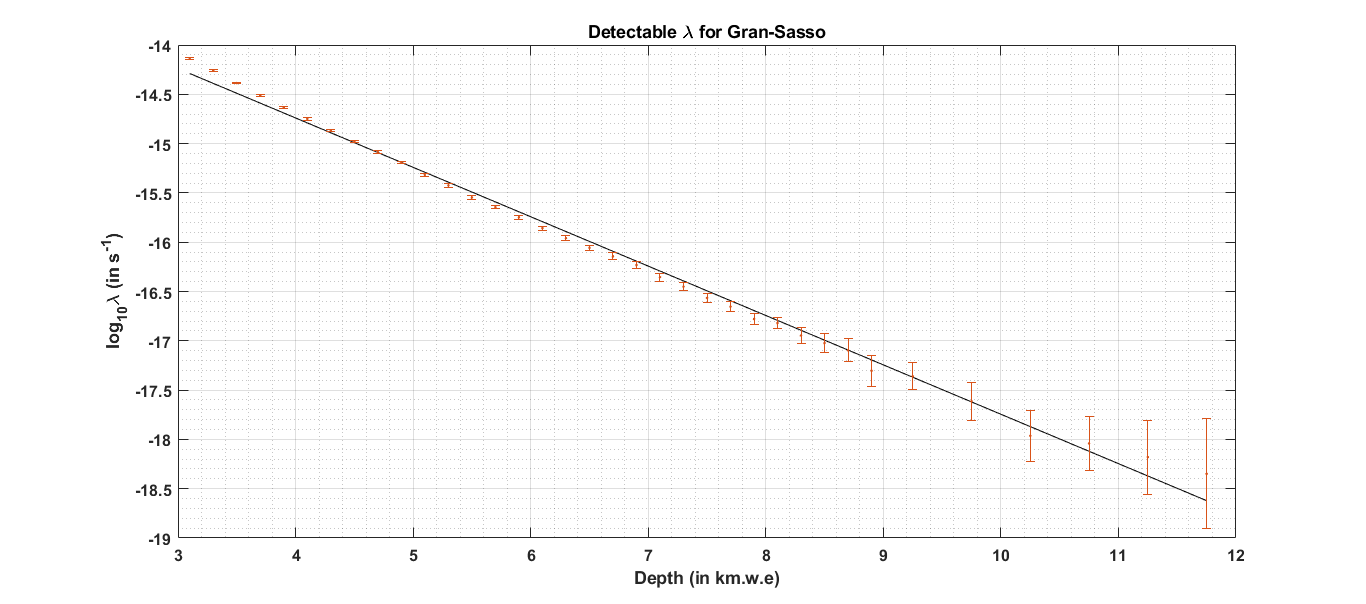}
	\caption{Variation of detectable $ \lambda $  with depth for  Gran-Sasso for a Ge cube of l=10 cm. For Gran-Sasso as $\log_{10}\lambda=-0.50d-12.74 $. Error in the intensity values \cite{e} have been taken into account. An error of $ 4\% $ has been assumed \cite{g} in the parameters for average energy taken from \cite{g} and \cite{h}.}
\end{figure}
\begin{figure}[H]
	\centering
	\includegraphics[width=1\linewidth]{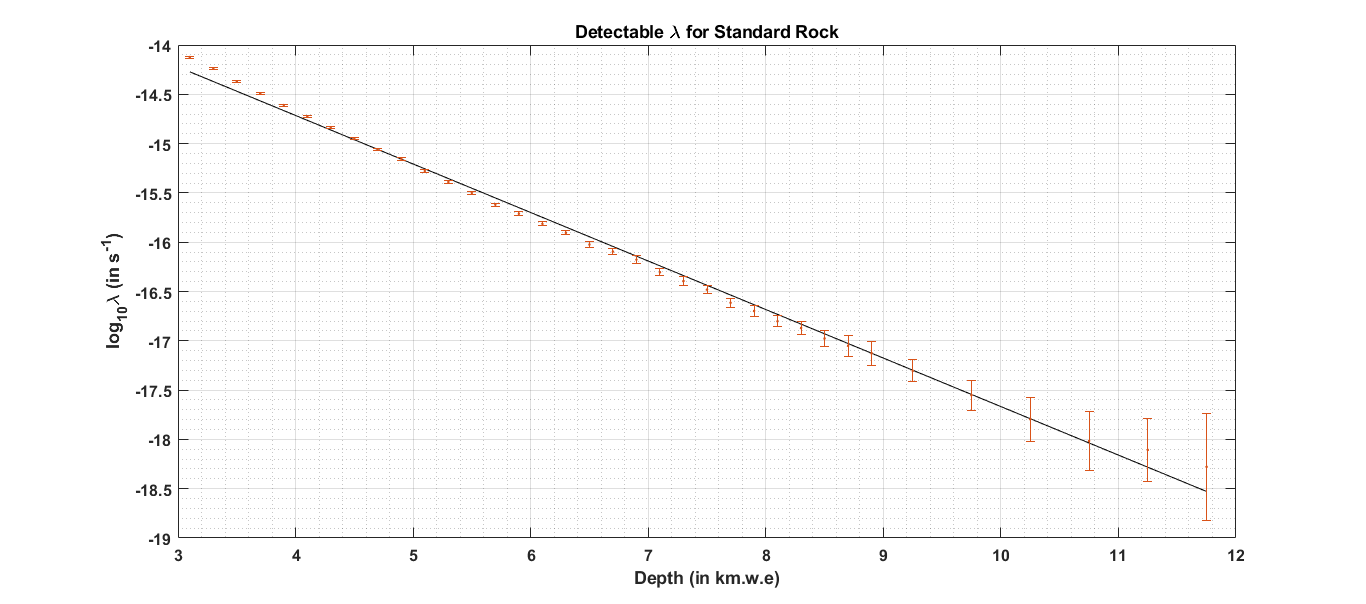}
	\caption{Variation of detectable $ \lambda $  with depth for Standard rock  for a Ge cube of l=10 cm. For standard rock it varies as $ \log_{10}\lambda=-0.49d-12.75 $. Error in the intensity values \cite{e} have been taken into account. An error of $ 4\% $ has been assumed \cite{g} in the parameters for average energy taken from \cite{g} and \cite{h}.}
\end{figure}

So, if gamma radiation and neutrons have been completely shielded out, going deeper can help  detect smaller values of $ \lambda $. In Fig. 9 $r_c$ is also allowed to vary, and $ \lambda $ vs $ \text{r}_c $ exclusion plot is shown. The contours show the detectable 
$\lambda, r_c$ values as a function of depth.
\begin{figure}[H]
	\centering
	\includegraphics[width=1.0\linewidth]{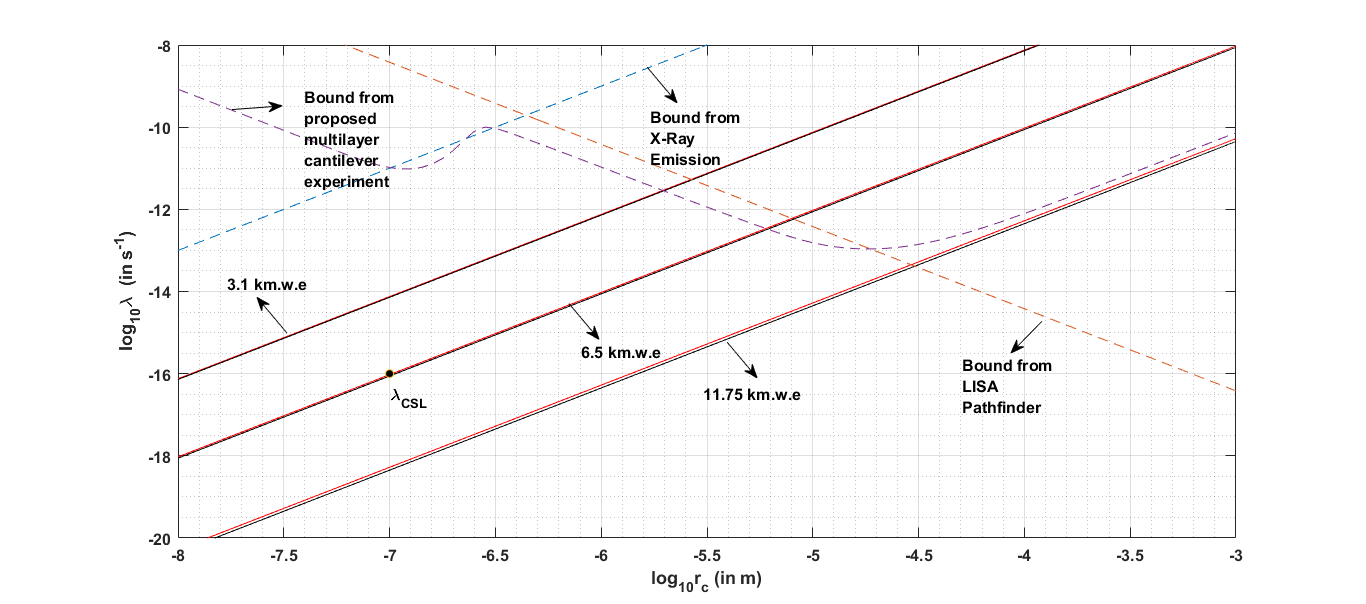}
	\caption{$ \lambda $ vs $ \text {r}_c $ plot for a Ge cube of l=10 cm. Contours for both Gran-Sasso and standard rock are plotted (Gran-Sasso in red and Standard rock in black). The depth is indicated for a pair of these lines. Other bounds as obtained from LISA Pathfinder \cite{carlessoLISA, LISA2}, X-ray spontaneous emission  \cite{curceanu} and that from proposed multilayer cantilever experiment \cite{carlesso} are also shown. 
	As shown, $\lambda_{CSL}$ is detectable at a depth of 6.5 km.w.e. These bounds are hypothetical, as no measurements of internal temperature were performed}
\end{figure}
Different lines are for different depths indicated by arrows. Some current bounds are also shown. For standard CSL values i.e. $ \lambda={10}^{-16}\text{s}^{-1}$ and $ r_c={10}^{-7}\text{m}^{-1} $, one would have to go to a depth of 6.3 km.w.e at Gran-Sasso and Standard rock. This is when power deposited due to cosmic ray muons is to be kept two orders of magnitude less than CSL heating. Currently, the deepest underground laboratory  is  the China Jinping Underground Laboratory, which is 6.7 km.w.e deep. It can  detect $ \lambda $ values up to $ 7.2\times{10}^{-17}\text{s}^{-1} $ assuming a difference of two orders of magnitude. Hence, $ \lambda_{CSL} $ can be detected with current technology. If we are to work with just one order of magnitude difference, the values for depth will improve, with 4.5 km.w.e at Gran-Sasso and  Standard rock. The proposed India-based Neutrino Observatory (INO) has a rock cover of 3.7 km.w.e. and can detect $\lambda\sim 10^{-14}$ s$^{-1}$.

\section{Conclusion}
The above calculations suggest that accounting for background noise for bulk heating experiments looking for CSL effects might not be a problem. In most cases, heating due to background noise can be completely neglected by going underground and using proper shielding. This will not only reduce the number of such background events but also the amount of power they deposit. The underground event rate would be actually very small, of the order of $10^{-7}$ s$^{-1}$ for a 10 cm detector at a depth of 6.5 km.w.e. [see Figs. 3, 4], so that it will  likely be possible to detect each individual event and subtract its effect.  

We end by noting that the (yet undetected) relic cosmic neutrino background is in principle a possible candidate for the CSL noise field \cite{RMP}. This background is thought to be nearly impossible to detect with present technology. What is interesting is that the cutoff frequency of relic neutrinos would be around $10^{11}$ Hz, so it should be on the edge of being in the right frequency range for a bulk heating experiment. This aspect is at present under further investigation.

 \bigskip
 
 {\bf Acknowledgement}: Ruchira Mishra would like to thank TIFR, Mumbai for hospitality during the Visiting Students Research Programme. A.V. acknowledges support from EU FET project TEQ (grant agreement 766900). A.V. thanks Stephen Adler for useful comments on an earlier version of the manuscript.
 



\end{document}